\documentstyle[12pt,titlepage]{article} 

\def\be{\begin{equation}}
\def\ee{\end{equation}}
\def\ben{\begin{eqnarray}}
\def\een{\end{eqnarray}}
\begin{document}
\setcounter{page}{1}

\thispagestyle{empty}
\noindent
\begin{center}
\large\bf{An $SU(2)_{L}\times U(1)_{Y} \times S_{3} \times D$ model \\ 
for atmospheric and solar neutrino deficits} \\ 
\vspace{10mm}
\normalsize{Asim K. Ray and Saswati Sarkar} \\ 
\vspace{5mm}
\normalsize{Department of Physics,Visva-Bharati University \\  Santiniketan-731235, India.} \\ 
\vspace{10mm}
\vspace{5mm}
\end{center}
\begin{abstract}

Motivated by the recent Super-Kamiokande experiment on atmospheric and solar                        
neutrino oscillation we propose a see-saw model of three generations 
of neutrinos based on the gauge group $SU(2)_{L}\times U(1)_{Y}$ with
discrete symmetries $(S_{3} \times D)$ and three right handed singlet neutrinos
so that this model can accommodate the recent Super-Kamiokande data on
atmospheric and solar neutrino oscillations. The model  
predicts maximal mixing between $\nu_{\mu}$ and $\nu_{\tau}$ with 
$sin^{2}2{\theta_{\mu\tau}}$ = 1 as required 
by the atmospheric neutrino data and small mixing between $\nu_{e}$ and 
$\nu_{\mu}$ with $sin^{2}2{\theta_{e\mu}} \sim (10^{-2}-10^{-3})$  
as a possible explanation of the solar neutrino deficit through
the MSW mechanism. The model admits two mass scales of which one breaks 
the electroweak symmetry and the other is responsible for the breaking of the
lepton number symmetry at GUT scale leading to small Majorana mass of the 
left handed doublet neutrinos.\\

PACS number(s).  12.60.Fr., 13.15.+g 13.40.Em., $\rightarrow$ 13.15.+g., 14.60.Pq.

\end{abstract}
\newpage

\section{Introduction}  

Recent results from the Super-Kamiokande experiment on atmospheric neutrino
anomaly [1] and solar neutrino deficit [2] have supported neutrino flavour
oscillation as possible explanation of these effects. Atmospheric neutrino 
data are consistent with 
$\nu_{\mu} \rightarrow \nu_{\tau}$ oscillations with $\Delta{ m^2}_{\mu\tau}
=(0.5 - 6) \times 10^{-3} eV^2$ and a nearly maximal 
mixing $sin^{2}2{\theta}_{\mu\tau}\ge 0.82$.
The data are equally consistent if $\nu_{\tau}$ is replaced by a sterile neutrino
and several authors have considered models with an extra light singlet neutrino, 
in additon to the usual three heavy right handed singlet neutrinos [3].

Furthermore, Super-Kamiokande data on solar neutrino deficit allow 
a solution in terms of $\nu_{e}$ disappearence vacuum oscillation [4] with 
$\Delta{ m^2}_{e\mu}
\sim 10^{-10} eV^2$ and a nearly maximal mixing angle $sin^{2}2{\theta}_{e\mu}
\sim 1.0$. Also consistent with these data are 
the small angle matter enhanced solution with 
$\Delta{ m^2}_{e\mu} \sim (0.5 - 1) \times 10^{-5} eV^2$ 
and $sin^{2}2{\theta}_{e\mu} \sim 10^{-2} - 10^{-3}$ [5]
as well as matter enhanced large angle solution with $\Delta{ m^2}_{e\mu} 
\sim (10^{-5} - 10^{-4}) eV^2$ and
$sin^{2}2{\theta}_{e\mu} \sim 1.0$ [6] although the small angle is more likely
at the moment. The recent CHOOZ [7] reactor result excludes the large mixing
angle neutrino oscillation of $\nu_{\mu} \rightarrow \nu_{e}$ as far as 
${\Delta{ m^2}_{\mu e}} \ge 9 \times 10^{-4} eV^{2}$.  
While the results of CHORUS [8] and NOMAD [9]
on $\nu_{\mu}$ oscillations are awaited and the LSND [10] data, which
are disfavoured by the KARMEN [11] experiment, needs confirmation 
from future experiments, several authors have proposed the strategy of
fitting the Super-Kamiokande data on atmospheric and solar neutrinos 
discarding the LSND data to discuss the problem of neutrino 
flavour oscillations within the framework of 
three generations of neutrinos. Such studies have shown that two mass squared
differences $\Delta m^{2}_{solar}$ and $\Delta m^{2}_{atmos}$ are relevant
and the hierarchy of light neutrino masses such as $ m_1, m_2 >> m_3$  or
$m_1, m_2 << m_3$ is adequate to explain the recent Super-Kamiokande data on 
neutrino oscillations. 
Another type of study [12] investigating the impact of neutrino 
oscillation on the texture of neutrino mass matrices in a model independent
way led to maximal mixing and appropriate mass squared difference, which are 
necessary to explain the Super-Kamiokande data.\\
Recently several authors [13,14] have 
studied see-saw models  based on the extension of the standard model 
including two singlet neutrinos with an 
extra $U(1)^{\prime}$ gauge group corresponding to a newly defined 
gauge charge , the gauge charge being $(B - 3L_{e})$ for Ref.[13] and 
$(B - 3/2 [L_{\mu} + L_{\tau}])$ for Ref.[14].
Assuming the $U(1)^{\prime}$ symmetry breaking scale to be of the order
of ($\sim 10^{12}-10^{16}$) GeV, the model naturally accounts for the large(small)
mixing solutions to the atmospheric (solar) neutrino oscillations and 
explains the hierarchy of the left handed light neutrino masses. This has motivated us 
to propose an $SU(2)_L \times U(1)_Y$ model with ($S_3 \times D$) discrete 
symmetries and three right handed singlet neutrinos to explain the results
of Super-Kamiokande experiment on atmospheric and solar neutrinos. 
Our model differs from those in Ref.[13] and Ref.[14] in not having extra 
gauge symmetry and in the choice of discrete symmetries.
The present model contains three right handed singlet neutrinos $\nu_{eR}, \nu_{\mu R},
\nu_{\tau R}$ along with three doublet Higgs fields $\phi_{1}, \phi_{2},
\phi_{3}$
and three singlet Higgs fields $\chi_{1}, \chi_{2}, \chi_{3}$. The doublet
Higgs fields are responsible for generating the Dirac mass terms of the 
neutrinos  where as right handed singlet  
neutrinos acquire their heavy masses through the singlet Higgs fields.
We have imposed an extra global symmetry in our model to start with to ensure 
lepton number conservation by making appropriate lepton number assignments
for the Higgs fields. In our model each singlet Higgs field is assigned with
a (-2) lepton number. As soon as $\chi_{1}, \chi_{2}, \chi_{3}$ develop their
non-zero vevs, the lepton number symmetry is broken spontaneously at the GUT scale
$(\sim 10^{16})$ GeV yielding three heavy right handed Majorana neutrino masses
$(M_{1}, M_{2}, M_{3})$ throuh the well-kwown see-saw mechanism, which 
induce small hierarchical masses to the left handed doublet 
neutrinos $(\nu_{e}, \nu_{\mu},
\nu_{\tau})$. The $(S_{3} \times D)$ discrete symmetry facilitates 
appropriate mixing 
between the left handed neutrino flavours in order to explain atmospheric
and solar neutrino data.\\

We organise our paper as follows. Section II contains the model. Section III
describes the Higgs potential, the neutrino mass matrix and its analysis.
Conclusions are presented in Section IV.

\newpage
\section{The Model} \label{model}    

In our present model the lepton and the Higgs fields have the following 
 representation contents under the gauge group $SU(2)_{L}\times U(1)_{Y}$
\ben
l_{iL} (2, -1)_{1},    e_{iR} (1, -2)_{1} , \nu_{eR} (1, 0)_{1}, 
\nu_{\mu R} (1, 0)_{1},   \nu_{\tau R} (1, 0)_{1} 
\een

where i = 1, 2, 3.

\ben                              
\phi_{i} (2, 1)_{0}, \chi_{i} (1, 0)_{-2} 
\nonumber\\ 
\een
The subscripts in the parenthesis correspond to the lepton number 
L $( = L_{e} + L_{\mu} + L_{\tau})$.

We consider the following VEVs of the Higgs fields:
\ben
<\phi> = \left(
\begin{array}{c}
<\phi_{i}^{+}>  \\
<\phi_{i}^{0}> \\
\end{array}\right)  =  \left(
\begin{array}{c}
0\\
v_{i}\\
\end{array}\right),\nonumber\\
<\chi^{0}_{i}> = k_{i}
\een

The following discrete symmetries have been incorporated in the model in 
order to generate the required pattern of Majorana neutrino mass matrix 
as well as the mixing: \\

(i) $S_{3}$- symmetry:   \\

\ben
l_{1L} \rightarrow 1, l_{2L} \rightarrow 1,  l_{3L} \rightarrow 1,   
(e_{R}, \tau_{R}) \rightarrow 2,  \mu_{R}, \rightarrow 1, 
(\nu_{eR}, \nu_{\tau R}) \rightarrow 2, \nu_{\mu R} \rightarrow 1,
\nonumber\\
(\phi_{1}, \phi_{3}) \rightarrow 2, \phi_{2} \rightarrow 1,
\chi_{1} \rightarrow 1, \chi_{2} \rightarrow 1,  \chi_{3} \rightarrow 1   
\een

(ii) D- symmetry:

\ben
l_{1L}\rightarrow i\omega^{*} l_{1L},                                
l_{2L}\rightarrow -i\omega^{*} l_{2L},  
l_{3L}\rightarrow -i\omega^{*} l_{3L},
e_{R} \rightarrow -i\omega^{*} e_{R},
\mu_{R}\rightarrow \mu_{R},\nonumber\\
\tau_{R}\rightarrow \omega^{*} \tau_{R},
\nu_{eR} \rightarrow -i\omega^{*} \nu_{eR},   
\nu_{\mu R} \rightarrow  \nu_{\mu R},
\nu_{\tau R} \rightarrow {\omega^{*}} \nu_{\tau R},\nonumber\\
\phi_{1}\rightarrow -i\phi_{1}, \phi_{2}\rightarrow i \omega\phi_{2}, 
\phi_{3}\rightarrow i\phi_{3}, 
\chi_{1} \rightarrow -{\omega}^{2} \chi_{1},
\chi_{2}\rightarrow \chi_{2}, 
\chi_{3}\rightarrow {\omega}^{2}\chi_{3}
\een

where $\omega = e^{2\pi i\over 3}$. \\

The choice of the discrete symmetries $S_{3} \times D$ allow $\nu_{eL}$
to couple with $\nu_{\tau R}$ only through $\phi_{1}$ Higgs field. But similar
couplings of $\nu_{eL}$ with other two singlet neutrinos are prohibited.
However, $\nu_{\mu L}$ and  $\nu_{\tau L}$ couple with the singlet
neutrinos through the Higgs fields $\phi_{2}$ and $\phi_{3}$. This
facilitates
small mixing between $\nu_{e}$ and $\nu_{\mu}$ as required for explanation of the
solar neutrino deficit and maximal mixing between $\nu_{\mu}$ and $\nu_{\tau}$
in order to explain atmospheric neutrino anomaly with appropriate 
choice of the vev's of the Higgs fields: 
$ <\phi_{1}> \sim 1 GeV $, $ <\phi_{2}>  =  <\phi_{3}> \sim 10^{2} GeV$.
The small vev of the Higgs field $\phi_{1}$ is achieved by choosing positive mass 
term of the $\phi_{1}$ Higgs field in the Higgs potential. The non-zero but small
vev of $\phi_{1}$ arises due to the presence of quartic coupling term 
in the Higgs potential [15].\\

\section{Higgs Potential and Neutrino Mass Matrix} \label{higgs}

We focus our attention on the relevant terms of the Higgs potential,
which yield small value of the vev of $\phi_{1}$ and, hence,
write explicitly the following $\phi_{1}$ dependent terms in the 
potential.

\ben
V_{\phi_{1}} = m_{1}^{2}(\phi_{1}^{\dagger}\phi_{1})
+ \lambda_{1}{(\phi_{1}^{\dagger}\phi_{1})}{(\phi_{2}^{\dagger}\phi_{2})}  
+ \lambda_{2}{(\phi_{1}^{\dagger}\phi_{1})}{(\phi_{3}^{\dagger}\phi_{3})}
+ \lambda_{3}{(\phi_{1}^{\dagger}\phi_{1})}{(\chi_{1}^{*}\chi_{1})}\nonumber\\
+ \lambda_{4}{(\phi_{1}^{\dagger}\phi_{1})}{(\chi_{2}^{*}\chi_{2})}  
+ \lambda_{5}{(\phi_{1}^{\dagger}\phi_{1})}{(\chi_{3}^{*}\chi_{3})}   
+  \lambda^{\prime} (\phi_{1}^{\dagger}\phi_{3}\chi_{1}^{*}\chi_{3} \nonumber\\
+ \chi_{3}^{*}\chi_{1}\phi_{3}^{\dagger}\phi_{1})
\een

After substituting the vev's of the Higgs fields in the 
above potential and minimising with respect to 
$<\phi_{1}> = v_{1}$, we obtain,

\ben
{v_{1}} = - {{\lambda^{\prime}v_{3}k_{1}k_{3}}\over {M_{\phi_1}^{2}}}
\een

where
\ben
{M_{\phi_1}^2} = m_{1}^{2} + \lambda_{1}v_{2}^{2} + \lambda_{2}v_{3}^{2} + 
\lambda_{3}{k_{1}^2} + \lambda_{4}k_{2}^2  
+ \lambda_{5}k_{3}^{2}
\een

is the physical mass of the $\phi_{1}$ Higgs field.

Assuming $m_{1}^{2}$ to be much greater 
than all other terms in Eqn.(8) and putting $M_{\phi_1}^{2} \sim m_{1}^{2}$,
which does not affect the essential 
results obtained in our model, we get,
\ben
v_{1} = {- {\lambda^{\prime}v_{3}k_{1}k_{3}}\over {m_1}^{2}} .
\een
We require the value of the vev of $<\phi_{1}> \sim 1 GeV$.
This would correspond to $\lambda^{\prime}\sim 1, v_{3} \sim 10^{2}$ GeV
$k_{1} \sim k_{3} $ and $m_{1} \sim 10 k_{3}$.

Turning to the neutrino sector, the most general discrete symmetry invariant
Yukawa interaction is given by
\ben
L_{y}^{\nu} = f_{1}\bar{l_{1L}}\nu_{\tau R}\tilde\phi_{1} 
+ f_{2}\bar{l_{2L}}\nu_{\mu R}\tilde\phi_{2}
+ f_{3}\bar{l_{2L}}\nu_{\tau R}\tilde\phi_{3}
+ f_{4}\bar{l_{3L}}\nu_{\mu R}\tilde\phi_{2}\nonumber\\
+ f_{1}\bar{l_{3L}}\nu_{\tau R}\tilde\phi_{3}
+ g_{1}\bar{\nu_{eR}^{c}}\nu_{eR} \chi_{1}^{0}
+ g_{2}\bar{\nu_{\mu R}^{c}}\nu_{\mu R} \chi_{2}^{0}  
+ g_{3}\bar{\nu_{\tau R}^{c}}\nu_{\tau R} \chi_{3}^{0} + h.c.  
\een

Substituting the vev's of the Higgs fields in Eqn.(10) we obtain the 
following form of (6x6) Majorana neutrino mass matrix in the basis \\
 $(\nu_{eL}, \nu_{\mu L}, \nu_{\tau L}, \nu_{eR}^{c}, \nu_{\mu R}^{c}, 
\nu_{\tau R}^{c})$:

\ben
M_{\nu} = \left(
\begin{array}{cc}
0 & D^T  \\
D & M \\
\end{array}\right)  = \left(\matrix{0& 0& 0& 0& 0&  c_{3}\cr
               0& 0& 0& 0& a_{2}& a_{3} \cr
               0& 0& 0& 0& b_{2}& b_{3} \cr
               0& 0& 0& M_{1}& 0 & 0 \cr
               0& a_{2}& b_{2}& 0& M_{2}& 0 \cr
               c_{3}& a_{3}& b_{3}& 0& 0& M_{3}}\right)
\een

where D and M denote the (3x3) Dirac and Majorana neutrino 
mass matrices respectively. The matrix elements of Eqn.(11) are given by

\ben
c_{3} = f_{1} v_{1}, a_{2} = f_{2} v_{2}, 
a_{3} = f_{3} v_{3}, b_{2} = f_{4} v_{2}, \nonumber\\
b_{3} = f_{5} v_{3}, M_{1} = g_{1} k_{1},  
M_{2} = g_{2} k_{2}, M_{3} = g_{3} k_{3}.
\een

We see from the (6x6) mass matrix that the right 
handed Majorana neutrino mass matrix M becomes diagonal with mass eigenvalues
$M_{1}$, $M_{2}$ and $M_{3}$ due to the imposition 
of the $(S_{3} \times D)$ discrete symmetries. 
While all the heavy right handed neutrinos
acquire their masses at the lepton number symmetry breaking scale which is of 
the order of $\sim 10^{16}$ GeV, we assume the following hierarchy of masses 

\ben
{M_{2}\over M_{3}} = {1\over{10}}
\een
in order to arrive at the required mass ratio for the left handed 
doublet neutrinos.

The induced (3x3) mass matrix for the left handed neutrinos $(\nu_{e}, \nu_{\mu},
\nu_{\tau})$ can be calculated from Eqn.(11) using the see-saw formula in the 
basis of a diagonal Majorana mass matrix as 
\ben
m_{ij} = {D_{1i}D_{1j}\over{ M_{1}}} + {D_{2i}D_{2j}\over{ M_{2}}} +
         {D_{3i}D_{3j}\over{ M_{3}}} 
\een
where i and j denote the three neutrino flavours. 
and $D_{1i},D_{2i} and D_{3i} $ refer to the respective Dirac mass matrix
elements in Eqn.(11). Thus we get

\ben
m_{\nu}=\left(\matrix{ {c_3^2}\over M_{3}& {c_{3}a_{3}}\over M_{3}& {c_{3}b_{3}}\over M_{3} \cr
               {c_{3}a_{3}}\over M_{3}& {{a_2^2}\over M_{2}} + {{a_3^2}\over M_{3}}& {{a_{2}b_{2}}\over M_{2}} + {{a_{3}b_{3}}\over M_{3}} \cr
               {c_{3}b_{3}}\over M_{3}& {{a_2 b_2}\over M_{2}} + {{a_3 b_3}\over M_{3}}& {{b_2^2}\over M_{2}} + {{b_3^2}\over M_{3}}}\right) 
\een

With the redefinition of the matrix elements as 

${{c_3}\over \sqrt {M_3}} \rightarrow C_{3}$,  ${{a_3}\over \sqrt {M_3}} \rightarrow A_{3}$, 
${{b_3}\over \sqrt {M_3}} \rightarrow B_{3}$,  ${{a_2}\over \sqrt {M_2}} \rightarrow A_{2}$, 
${{b_2}\over \sqrt {M_2}} \rightarrow B_{2}$ \\

Eqn.(15) is reduced to the following simple form 

\ben
m_{\nu}=\left(\matrix{ {C_3^2}& {C_{3}A_{3}} & {C_{3}B_{3}} \cr
               {C_{3}A_{3}} & {A_2^2} + {A_3^2}& {A_{2}B_{2}}+{A_{3}B_{3}}\cr
               {C_{3}B_{3}}& {A_{2}B_{2}}+{A_{3}B_{3}}& {B_2^2}+{B_3^2}}\right) 
\een

We now proceed to calculate the masses and mixing angles of the three light
left handed neutrinos by diagonalising the mass matrix in Eqn.(16). It can be
shown that the determinant of $m_{\nu}$ is zero implying that
one of the eigen values of $m_{\nu}$ is zero.\\

We then assume the approximation,$ A_2, B_2 >> A_3, B_3 >> C_3 $, which will be 
true for our parameter space of interest. After diagonalisation of the (3x3) 
induced neutrino mass matrix we get the following mass eigen values:

\ben
m_1 \sim {A_2^2 + B_2^2}\nonumber\\
m_2 \sim {{( A_{2} B_{3} - A_{3} B_{2})}^{2}\over { A_{2}^{2} + B_{2}^{2}}} \nonumber\\
m_3 = 0
\een

where ${m_{2}\over m_{1}} {\sim {M_{2}\over M_{3}}}$.

Finally, we investigate the mixing matrix connecting the three flavour
eigenstates $(\nu_{e}, \nu_{\mu}, \nu_{\tau})$ to the mass eigen states
$(\nu_{3}, \nu_{2}, \nu_{1})$ in increasing order of mass:

\ben
 \left(
\begin{array}{c}
\nu_{e} \\
\nu_{\mu} \\
\nu_{\tau}\\
\end{array}\right)  =  \left(
\begin{array}{ccc}
1  & { -C_{3} \sqrt {(A_{2}^{2}+B_{2}^{2})}} \over {A_{2}B_{3} - A_{3}B_{2}} & -1 \\
{C_{3}B_{2}}\over {A_{2}B_{3} - A_{3}B_{2}} &  B_{2}\over {\sqrt {A_{2}^{2}+B_{2}^{2}}} & {C_{3}A_{2}}\over {A_{2}A_{3} + B_{2}B_{3}}   \\
{-C_{3}A_{2}}\over {A_{2}B_{3} - A_{3}B_{2}} & -A_{2}\over {\sqrt {A_{2}^{2}+B_{2}^{2}}} & {C_{3}B_{2}}\over {A_{2}A_{3} + B_{2}B_{3}}   \\   
\end{array}\right)  \left(        
\begin{array}{c}
\nu_{3}\\
\nu_{2}\\
\nu_{1}\\
\end{array}\right)\nonumber\\
\een

In our model the mixing angle between $\nu_{\mu}$ and  $\nu_{\tau}$ is
given by ratio of the two Yukawa couplings as follows:

\ben
tan \theta_{\mu \tau} \sim {A_{2}\over B_{2}} \sim {f_{2}\over f_{4}}
\een
Assuming these Yukawa couplings to be equal, we obtain
$\theta_{\mu\tau}  = 45^{0}$ ie, $sin^{2}{2 \theta_{\mu\tau}} = 1$. This
agrees with the Super-Kamiokande data on atmosoheric neutrinos, which suggest
maximal mixing between $\nu_{\mu}$ and $\nu_{\tau}$.
Now, for the solar neutrino oscillation $\nu_{e} \rightarrow \nu_{\mu}$,
the mixing angle between $\nu_{e} $ and $\nu_{\mu}$ in our model is given by

\ben
sin \theta_{e \mu} \sim {C_{3} \sqrt {(A_{2}^{2}+B_{2}^{2})} 
\over {A_2 B_3 - A_3 B_2}} \sim{ f_{1} v_{1} \over f_{3,5} v_{3}}
\een
If the couplings are assumed to be of the same order, we get

\ben
sin \theta_{e \mu} \sim {v_{1} \over v_{3}} {\sim 10^{-2}}.
\een
This is in good agreement with the recent Super-Kamiokande solar neutrino data.\\

We shall now estimate the two nonzero light neutrino masses
in our model. Assuming  $v_{3} = 1.5 \times 10^{2}$ GeV and $M_{2} \sim 10^{15} $GeV
and $M_{3} \sim 10^{16}$ GeV, we get $m_{\nu_{\mu}} \sim 0.0023$ eV and
$m_{\nu_{\tau}} \sim 0.045$ eV. Thus our model leads to  
$\Delta m_{\mu \tau}^{2} \sim 2 \times 10^{-3} eV^{2}$ and
$\Delta m_{e \mu}^{2} \sim 0.53 \times 10^{-5} eV^{2}$, which are consistent
with the experimental results of Super-Kamiokande.

\section{Conclusions} \label{Conclusion}

In the present paper we have focused our attention on the Super-Kamiokande 
data on atmospheric and solar 
neutrinos without taking into account the LSND result.   
We demonstrate that the evidence of atmospheric and solar neutrino 
oscillations provided by Super-Kamiokande experiments can be accommodated
in an extension of the standard electroweak gauge model based on the gauge group
$SU(2)_{L} \times U(1)_{Y}$ with appropriate discrete symmetries $(S_{3} 
\times D)$ and Higgs fields along with the three right handed singlet 
neutrinos. The model can simultaneously reconcile
the maximal mixing between $\nu_{\mu}$ and $\nu_{\tau}$ as required by the 
atmospheric neutrino data as well as matter enhanced MSW solution to solar neutrino
problem with small mixing angle between $\nu_{e}$ and $\nu_{\mu}$.
The desired small masses of the left handed neutrinos are generated by the 
well known see-saw mechanism with reasonable choice of the model parameters.
Apart from the electroweak symmetry breaking scale, the model admits a 
lepton number symmetry breaking scale at $\sim 10^{16}$ GeV.

\section*{Acknowledgements}

S.S. acknowledges the financial support provided by the CSIR in
the form of a Research Associateship. A.K.R. thanks D. P. Roy 
for many helpful discussions.\\


\newpage
\begin{thebibliography}{99}
\bibitem{kam} Super-Kamiokande Collaboration: Y. Fukuda et al, 
Phys. Rev. Lett. 81, 1562 (1998); hep-ex/9805006; Phys Lett B433, 9 (1998); 
T. Kajita, Talk presented at Neutrino 98, Takayama, Japan (1998).
\bibitem{sup} Super-Kamiokande Collaboration: Y. Fukuda et al, 
Phys. Rev. Lett. 81, 1158 (1998); Y. Suzuki, 
Talk presented at Neutrino 98, Takayama, Japan (1998).
\bibitem{ema} E. Ma, P. Roy, Phys. Rev. D52, R4780 (1995); N. Gaur, A. Ghosal,
E. Ma, P.Roy, UCRHEP-T233, MRI-PHY/980444, (1998).
\bibitem{bar} V. Barger, R. J. N. Phillips and K. Whisnant, Phys Rev. D24,
538 (1981); S. L. Glashow and L. M. Krauss, Phys. Lett. B190, 199 (1987);
A. Acker, S. Pakvasa and J. Pantaleone, Phys. Rev. D43, 1754 (1991);
P. I. Krastev and S. T. Petkov, Phys. Lett. B285, 85 (1992); Phys. Rev. Lett.
72, 1960 (1994); Phys. Rev. D53, 1665 (1996).
\bibitem{hat} N. Hata and P. G. Langacker, Phys. Rev. D56, 6107 (1997).
\bibitem{bah} J. N. Bahcall, P. J. Krastev and A. Yu. Smirnov, hep-ph/9807216.  
\bibitem{chz} M. Appollonio et al Phys. Lett B420, 397 (1998).
\bibitem{cho} E. Eskut et al., CHORUS Collaboration, Phys. Lett. B424, 202 (1998).
\bibitem{nom} J. Altegoer et al., NOMAD Collaboration, CERN-EP/98-57.
\bibitem{lsn} C.Anthassopoulas et. al.,(LSND Collaboration), Phys. Rev. 
Lett. 75,2650 (1995); Phys Rev. C54, 2685 (1996); Phys. Rev. Lett. 77, 3082 (1996); 
nucl-ex/9706006; nucl-ex/9709006; Phys. Rev. Lett. 81, 1774 (1998).
\bibitem{kar} KARMEN Collaboration: R. Armbruster et al., Phys. Rev. C57, 3414
(1998); Phys. Lett. B423, 15 (1998); K. Eitel et al., hep-ex/9809007.
\bibitem{alt} G. Altarelli and F. Feruglio, Phys. Lett. B439, 112 (1998);
M. Bando, T. Kugo and K. Yoshioka, Phys. Rev. Lett 80, 3004 (1998).
\bibitem{usa} E. Ma, D.P. Roy and U. Sarkar, UCRHEP-T236, (1998). 
\bibitem{emp} E. Ma and  D. P. Roy, DTP/98/78; UCRHEP-T210, (1998). 
\bibitem{asm} A. K. Ray and S. Sarkar, Phys. Rev. D58, 055010 (1998).
\end{thebibliography}
\end{document}